\documentclass[prl,twocolumn,showpacs]{revtex4}
\usepackage{graphicx}
\usepackage{graphics}
\usepackage{amsfonts}
\usepackage{amsmath}
\usepackage{epsfig}

\begin{document}
\catcode`\ä = \active \catcode`\ö = \active \catcode`\ü = \active
\catcode`\Ä = \active \catcode`\Ö = \active \catcode`\Ü = \active
\catcode`\ß = \active \catcode`\é = \active \catcode`\è = \active
\catcode`\ë = \active \catcode`\ô = \active \catcode`\ê = \active
\catcode`\ø = \active \catcode`\ò = \active \catcode`\í = \active
\catcode`\Ó = \active \catcode`\ú = \active \catcode`\á = \active
\catcode`\ã = \active
\defä{\"a} \defö{\"o} \defü{\"u} \defÄ{\"A} \defÖ{\"O} \defÜ{\"U} \defß{\ss} \defé{\'{e}}
\defè{\`{e}} \defë{\"{e}} \defô{\^{o}} \defê{\^{e}} \defø{\o} \defò{\`{o}} \defí{\'{i}}
\defÓ{\'{O}} \defú{\'{u}} \defá{\'{a}} \defã{\~{a}}



\newcommand{\li}{$^6$Li}
\newcommand{\na}{$^{23}$Na}
\newcommand{\cs}{$^{133}$Cs}
\newcommand{\kk}{$^{40}$K}
\newcommand{\rb}{$^{87}$Rb}
\newcommand{\vect}[1]{\mathbf #1}
\newcommand{\g}{g^{(2)}}
\newcommand{\one}{$\left|1\right>$}
\newcommand{\two}{$\left|2\right>$}
\newcommand{\V}{V_{12}}
\newcommand{\kfa}{\frac{1}{k_F a}}

\title{Formation Time of a Fermion Pair Condensate}

\author{M.W. Zwierlein, C.H. Schunck, C.A. Stan, S.M.F. Raupach, and W. Ketterle}

\affiliation{Department of Physics\mbox{,} MIT-Harvard Center for
Ultracold Atoms\mbox{,} and Research Laboratory of Electronics,\\
MIT, Cambridge, MA 02139}

\date{\today}

\begin{abstract}

The formation time of a condensate of fermionic atom pairs close to a Feshbach resonance was studied.
This was done using a phase-shift method in which the
delayed response of the many-body system to a modulation of the
interaction strength was recorded. The observable was the fraction of
condensed molecules in the cloud after a rapid magnetic field ramp
across the Feshbach resonance. The measured response time was slow
compared to the rapid ramp, which provides final proof that the
molecular condensates reflect the presence of fermion pair
condensates before the ramp.
\end{abstract}

\maketitle

Atomic Fermi gases close to a Feshbach resonance~\cite{ties93,stwa76,inou98,cour98fesh} offer the
unique possibility of studying many-body phenomena in a strongly interacting system with tunable interactions.
Recently a major focus has been on condensates of pairs of fermionic atoms~\cite{grei03mol_bec,joch03bec,zwie03molBEC,bour04coll,rega04,zwie04rescond,kina04sfluid}.
By changing the magnetic field the interaction strength between atoms
in two spin states can be varied. That way, condensates of either
tightly bound molecules or of extended pairs of fermions can be
created, whose size can become comparable or even larger than the
interparticle spacing. The description of this so-called BEC-BCS
crossover~\cite{eagl69,legg80,nozi85} is an active frontier in many-body physics with still controversial interpretations~\cite{falc04,Bara04Coex,Ho04proj,Simo04}.

The control of interactions via magnetic fields does not only give
access to very different physical regimes, it also allows to apply a
time-varying interaction strength ~\cite{matt98,grei04spec} and to
study the dynamics of a many-body system in novel ways. This was
used in recent experiments in which molecular condensates were
observed after a rapid field ramp from the BCS to the BEC side of
the Feshbach resonance~\cite{rega04,zwie04rescond}. It was argued
that if the ramp time was faster than the formation time of a
molecular condensate, its presence after the sweep necessarily
reflected a preexisting condensate of fermion pairs. However, without
access to that formation time, secondary evidence was gathered,
namely the invariance of the condensate fraction under variations of
the sweep rate~\cite{rega04} or of the density immediately before the
ramp~\cite{zwie04rescond}. This excluded simple models of the
molecular condensate formation during the ramp, but left room for
more sophisticated many-body effects. In
particular, the time to cross the Feshbach resonance in these
experiments was not faster than the unitarity limited
collision time $\propto \hbar E_F^{-1}$, and therefore dynamics
during the sweep could not be ruled out.

Here we present an experimental study of the formation time of a
fermionic condensate on the BCS side of the Feshbach resonance~\cite{foot:workshop}. We
employ a novel phase-shift method, which records the delayed response
of the many-body system to a modulation of the magnetic field that changes periodically its interaction strength. The observable is again the molecular
condensate fraction after a rapid sweep to the BEC side of the
Feshbach resonance. Its sensitivity to changes in the scattering
length on the BCS side~\cite{rega04,zwie04rescond} arises through the
dependence of the critical temperature for pair condensation on the
interaction strength. By showing that the delayed response time of
the molecular condensate fraction is long compared to the sweep times
used in the present and previous experiments, we infer that the
observed condensates could not have been created during the rapid
transfer and thus must originate from pre-existing fermion pair
condensates. However, we do find evidence that
condensed pairs are more likely to be transferred into molecules than
thermal pairs. Therefore, in contrast to assumptions made in previous
work~\cite{rega04, zwie04rescond}, the molecular condensate fraction
after the ramp does not equal the fraction of condensed atom
pairs above resonance.

The experimental setup was the same as in our previous work~\cite{zwie04rescond}.
A degenerate cloud of \li, sympathetically cooled with \na, 
was loaded into an optical dipole trap to access a broad
Feshbach resonance at 834 G~\cite{schu04fesh,bart04fesh} between the
two lowest hyperfine states of \li, labelled \one\ and \two. An equal
mixture of these states was evaporatively cooled at 770 G using an
exponential ramp-down of the optical trap to 15 mW. This resulted in
an essentially pure Bose-Einstein condensate of $3\cdot 10^6$
molecules. An upper limit for the temperature of the gas is
$\frac{T}{T_F} < 0.2$, with the Fermi temperature $T_F$ given by the zero-temperature, ideal gas
relation $T_F = \hbar \omega (3 N)^{1/3}$, $\omega/2\pi$ is the geometric mean of the trapping frequencies, and $N$ the total atom number. Next, the trap was recompressed to 25 mW
(trap frequencies: $\nu_x = \nu_y = 580 \,\rm Hz$, $\nu_z = 12.1 \,{\rm
Hz}\sqrt{0.2+B}$ with the magnetic field $B$ in kG.) and the magnetic
field was adiabatically increased in 500 ms to 1000 G, the starting
point for the following experiments. Here, in the wings of the
Feshbach resonance, the scattering length $a$ was still sufficiently
large and negative for the gas to be in the strongly interacting
regime, with $k_F \left|a\right| = 1.6$ at a Fermi energy of $E_F =
2.0\, \mu \rm K$ and a Fermi wavenumber $k_F = 1 / 2700 \; a_0$. The
temperature at this point could therefore not be reliably determined,
but is expected to be significantly lower than the one on the
BEC side due to adiabatic cooling~\cite{carr03}. Subsequently, the
magnetic field and thus the interaction strength in the gas were
modulated at frequencies in the range of 100 Hz - 500 Hz, and an
amplitude of about 50 G~\cite{foot:fieldcalib}. At a variable time $t$ after the start of
the modulation, the fraction of condensed fermion pairs was recorded
by time-of-flight analysis.

\begin{figure}
    \begin{center}
    \includegraphics[width=3.3in]{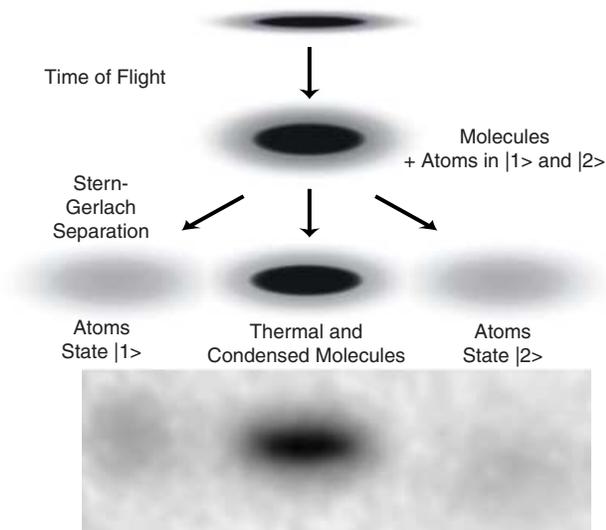}
    \caption[Title]{Imaging of molecular condensates. The rapid ramp
to zero field after release from the trap created a cloud containing both molecules and unpaired
atoms. A Stern-Gerlach field gradient separated atoms (magnetic moment $\pm
\frac{1}{3}\mu_B$ for state $\left|1\right>$ and $\left|2\right>$, resp.) from molecules, which are purely singlet at zero field.
At the end of 5 ms of ballistic expansion,
the molecules were dissociated in a fast ramp (in 3 ms to $\sim$ 1200
G) across the Feshbach resonance. After another 2 ms expansion again
at zero field, an absorption image of the separated clouds was taken. Condensate fractions
were determined from the molecular cloud, and the numbers in each
component were recorded. An absorption image is shown on the
bottom, the field of view is 3 mm x 1 mm.} \label{fig:Imaging}
    \end{center}
\end{figure}

To identify fermionic condensates across the resonance region, we
proceeded as in~\cite{rega04,zwie04rescond}. Immediately after
release of the cloud from the optical potential, the magnetic field
was switched to zero field (initial ramp-rate 30 G/$\mu$s), where
further expansion of the cloud took place. This rapid ramp out of the
resonance region transformed large fermion pairs into deeply bound
molecules with high efficiency~\cite{foot:noatoms}. Fig.~\ref{fig:Imaging} details the imaging procedure used to
determine molecular condensate fractions and the number of unpaired
atoms in each state after the ramp. In our previous work, we showed that the condensate fractions had a peak around the Feshbach resonance and
fell off on either side~\cite{zwie04rescond}. Here, this dependence was
exploited to observe the delayed response of the system to the
magnetic field modulation on the BCS side.

\begin{figure}
    \begin{center}
    \includegraphics[width=3.3in]{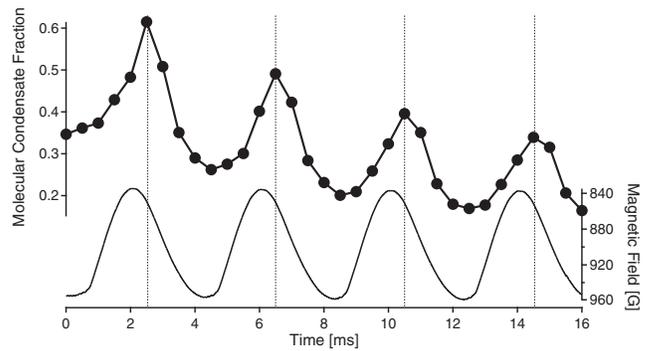}
    \caption[Title]{Measurement of the formation time of fermionic
pair condensates. Shown is the delayed response of the observed
condensate fraction (data points and thick line to guide the eye) to
a 250 Hz magnetic field modulation (thin line) on the BCS side of the
Feshbach resonance at 834 G. The condensates were detected as
described in Fig.~\ref{fig:Imaging}. Three measurements per point were taken in random order, the size of the data points reflecting the standard deviation. The vertical lines indicate the points of maximum condensate fraction,
which are delayed with respect to the times at which the magnetic field is closest to resonance.} \label{fig:delay}
    \end{center}
\end{figure}

Fig.~\ref{fig:delay} shows the main result of this paper:
The condensate fraction in the molecular clouds
after the rapid ramp did not follow the magnetic field modulation instantaneously, but lagged
behind. At a
Fermi energy of $E_F = 2 \,\mu \rm K$, the peak condensate fraction was delayed by
$\tau_R = (500 \pm 100) \,\mu$s with respect to the magnetic field's closest approach to resonance~\cite{foot:evaporation}. This timescale was independent of the modulation
frequency (compare Fig.~\ref{fig:delay} and Fig.~\ref{fig:correlation}a). $\tau_R$ equals 130 times the
unitarity limited collision time, $\hbar E_F^{-1} = 3.8 \,\mu s$. The
rapid magnetic field ramp utilized here and in~\cite{zwie04rescond}
traversed the Feshbach resonance in less than 10 $\mu {\rm s}$, which
is much smaller than $\tau_R$. 

This delay time can be interpreted as the relaxation time of the fermionic condensate.
In a normal Fermi gas of $N$ particles at temperatures much smaller
than the Fermi temperature $T_F$, relaxation occurs through
collisions between the thermally excited particles close to the Fermi
surface, whose number scales as $N_{\rm th} \simeq N \frac{k_B T}{E_F}$. The
number of available scattering states again being proportional to $\frac{k_B T}{E_F}$, the relaxation time will be $\tau_R \simeq \hbar \frac{E_F}{(k_B T)^2}$. In general, if the Fermi surface is smeared
out over an energy width $\Delta E$, the relaxation time is $\simeq
\hbar \frac{E_F}{(\Delta E)^2}$. This formula with $\Delta E=\Delta$
should apply also to the superfluid state~\cite{bara04Rabi} when the gap
parameter $\Delta$ is rapidly changed to a much smaller value.
Generally, one would expect $\Delta E$ to be the larger of $\Delta$
and $k_B T$.  Using $\tau_R = 500 \,\mu$s, we obtain the estimate
$\Delta E = 0.1 E_F$ which may set an upper bound for both
temperature and pairing gap.

The observed decrease in condensate fraction for subsequent cycles of
the magnetic field modulation could be due to heating.  Heating could
be caused by excitation of the cloud via the small accompanying
variation of the magnetic field curvature, by the frequent crossing
of the phase transition in low-density regions of the cloud, or by
repeatedly crossing the point of hydrodynamic
breakdown~\cite{bart04coll,kina04hydr}, where the pairing gap becomes
comparable to radial collective mode energies $\simeq 2 \hbar
\omega_r$.

In a compressed trap of $p=150 \rm mW$, at a 1.8 times higher Fermi
energy of $3.6 \,\mu \rm K$, the measured delay time was $\tau_R \simeq
(230 \pm 100) \,\mu s$.  BCS theory predicts that the relaxation time
should scale with density like $\tau_R \propto E_F^{-1}
e^{\frac{\pi}{k_F \left|a\right|}}$, giving
$\tau_R \simeq 200\,\mu\rm s$ for this experiment performed around 900 G. However, we regard this agreement with
observation as fortuitous since BCS theory cannot be rigorously
applied, and finite temperature effects may contribute to the
relaxation.

\begin{figure}
    \begin{center}
    \includegraphics[width=3.3in]{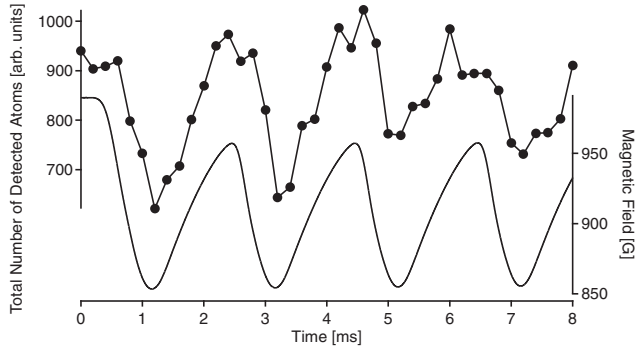}
    \caption[Title]{Total number of detected atoms (unbound atoms and molecules) after the
rapid ramp (same data set as in Fig.~\ref{fig:correlation}). It is modulated in phase with the magnetic field. For initial fields close to resonance, more atoms are "missing" after the rapid ramp.} \label{fig:totalsignal}
    \end{center}
\end{figure}

We now discuss further observations regarding the efficiency of
converting atoms into molecules.  Since the relaxation time
introduces some hysteresis, we observe the same condensate fraction
at two different magnetic field values.  Therefore, in contrast to
equilibrium experiments~\cite{rega04, zwie04rescond}, we can distinguish the
dependence of the conversion efficiency on condensate fraction and
magnetic field.

Fig.~\ref{fig:totalsignal} shows that the total number of detected atoms (in both the
atom and molecule channels) was modulated by the magnetic field. We
assume that this instantaneous response reflects the two-body physics
during the magnetic field sweep. In a simple two-state Landau-Zener
model, the initial magnetic field and the sweep rate determine what
fraction of the atoms appears as bound molecules.  However, the total
number of bound or unbound atoms should be constant in contrast to our
observations. This is evidence for the presence of other molecular
states which are populated during the magnetic field sweep, and the
population is larger for initial magnetic fields closer to the
Feshbach resonance. Note that the determination of the condensate
fraction is immune against those ``disappeared'' molecules, since the
two-body physics does not depend on the center-of-mass motion of the
atom pair.

\begin{figure}
    \begin{center}
    \includegraphics[width=3.3in]{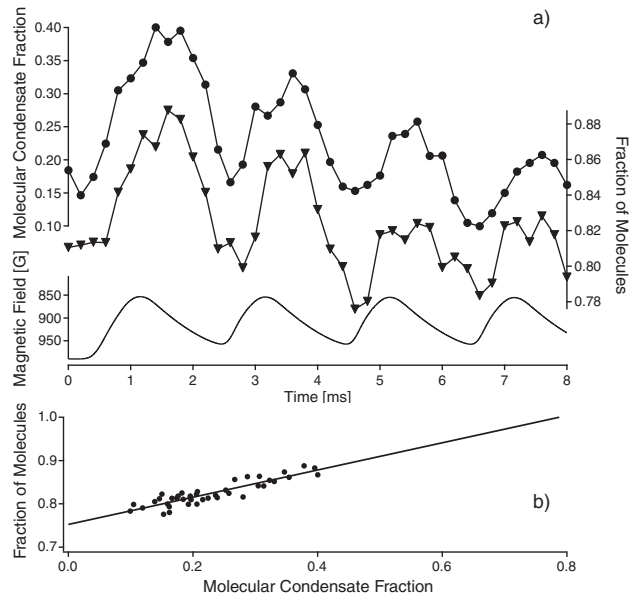}
    \caption[Title]{Correlation between the observed condensate
fraction and the molecular fraction. Shown are a) the condensate fraction vs time during a 500 Hz field modulation (circles), the fraction of molecules (triangles) and the magnetic field. Unlike the total detected signal (Fig.~\ref{fig:totalsignal}), the molecular fraction is modulated not in
phase with the magnetic field, but in complete correlation with the condensate fraction. Fig. b) displays the atomic signal vs condensate fraction, together
with a fitted line through the data.} \label{fig:correlation}
    \end{center}
\end{figure}

We now look at the molecular fraction which we define as $1 -
N_{\rm atom}/N_{\rm total}$, where $N_{\rm atom}$ is the number of atoms observed
after the sweep and $N_{\rm total}$ the total number of atoms before the
sweep (this definition includes the disappeared molecules). If the
molecule fraction would follow the instantaneous magnetic field, it
would again reflect the two-body physics during the sweep.  Instead,
we observe a delayed response in perfect correlation with the
condensate fraction  (Fig.~\ref{fig:correlation}).  Since the
delay time reflects the many-body physics of condensate formation,
this is clear evidence that the molecule conversion efficiency
depends on the initial many-body state.

The results show that fermion pairs occupying the zero-momentum
state are more completely transferred into tightly bound molecules
than thermal pairs. Extrapolating the fitted line in
Fig.~\ref{fig:correlation} b) to zero condensate fraction gives
the transfer efficiency from thermal atom pairs to molecules
(including the missing fraction) as $p_{\rm th} = 75\%$~\cite{foot:noatoms}. Extrapolating
towards the other limit, we do not expect any unpaired atoms after
the ramp already for a condensate fraction of 80\% \cite{foot:pure}, giving a transfer efficiency for condensed
fermion pairs into molecules of $p_0 = 100\%$. This effect leads to
an overestimate of the fermionic condensate fraction before the
sweep. Small condensate fractions will be overestimated by as much as
$\frac{p_0 - p_{\rm th}}{p_{\rm th}} = 33\%$. The largest absolute
error occurs for an initial pair condensate fraction of
$\frac{\sqrt{p_{\rm th}}}{\sqrt{p_0}+\sqrt{p_{\rm th}}} = 46\%$ and is about
$7\%$ in our case.

This effect has several possible explanations:  One is that the atomic
separation in a condensed atom pair is smaller than that of two
uncondensed atoms.  Also, the presence
of a large pair condensate increases the density of the cloud~\cite{foot:overestimate}.
Finally, if there are incoherent processes involved during the rapid
ramp, bosonic stimulation into the molecular condensate could play 
role.

In conclusion, we have determined the intrinsic timescale for the
growth of a fermion pair condensate by observing the delayed response
of the system to a change in its interaction strength. For our trap
parameters, the response was delayed by $\approx 500 \,\mu s$.  This time is far
longer than the time spent within the resonance region during the
conversion of fermion pairs into molecules. This provides final
proof that the observed molecular condensates originated from
condensates of pairs of fermions above the resonance. Regarding the
two-body physics of the rapid transfer, we found that there is a
missing fraction of particles after the ramp, presumably transferred
into unobserved molecular states. We found evidence that condensed
fermion pairs are more efficiently transformed into molecules than
thermal pairs during the rapid ramp. Thus, the observed
molecular condensate fractions tend to overestimate the initial
fermion pair condensate fraction.

This work was supported by the NSF, ONR, ARO, and NASA. We would
like to thank Michele Saba for the critical reading of the manuscript. S.\ Raupach is grateful to the Dr.
J\"urgen Ulderup foundation for financial support.


\begin{thebibliography}{32}
\expandafter\ifx\csname natexlab\endcsname\relax\def\natexlab#1{#1}\fi
\expandafter\ifx\csname bibnamefont\endcsname\relax
  \def\bibnamefont#1{#1}\fi
\expandafter\ifx\csname bibfnamefont\endcsname\relax
  \def\bibfnamefont#1{#1}\fi
\expandafter\ifx\csname citenamefont\endcsname\relax
  \def\citenamefont#1{#1}\fi
\expandafter\ifx\csname url\endcsname\relax
  \def\url#1{\texttt{#1}}\fi
\expandafter\ifx\csname urlprefix\endcsname\relax\def\urlprefix{URL }\fi
\providecommand{\bibinfo}[2]{#2}
\providecommand{\eprint}[2][]{\url{#2}}

\bibitem[{\citenamefont{Tiesinga et~al.}(1993)\citenamefont{Tiesinga, Verhaar,
  and Stoof}}]{ties93}
\bibinfo{author}{\bibfnamefont{E.}~\bibnamefont{Tiesinga}},
  \bibinfo{author}{\bibfnamefont{B.~J.} \bibnamefont{Verhaar}},
  \bibnamefont{and} \bibinfo{author}{\bibfnamefont{H.~T.~C.}
  \bibnamefont{Stoof}}, \bibinfo{journal}{Phys. Rev. A}
  \textbf{\bibinfo{volume}{47}}, \bibinfo{pages}{4114} (\bibinfo{year}{1993}).

\bibitem[{\citenamefont{Stwalley}(1976)}]{stwa76}
\bibinfo{author}{\bibfnamefont{W.~C.} \bibnamefont{Stwalley}},
  \bibinfo{journal}{Phys. Rev. Lett.} \textbf{\bibinfo{volume}{37}},
  \bibinfo{pages}{1628} (\bibinfo{year}{1976}).

\bibitem[{\citenamefont{Inouye et~al.}(1998)\citenamefont{Inouye, Andrews,
  Stenger, Miesner, Stamper-Kurn, and Ketterle}}]{inou98}
\bibinfo{author}{\bibfnamefont{S.}~\bibnamefont{Inouye}},
  \bibinfo{author}{\bibfnamefont{M.~R.} \bibnamefont{Andrews}},
  \bibinfo{author}{\bibfnamefont{J.}~\bibnamefont{Stenger}},
  \bibinfo{author}{\bibfnamefont{H.-J.} \bibnamefont{Miesner}},
  \bibinfo{author}{\bibfnamefont{D.~M.} \bibnamefont{Stamper-Kurn}},
  \bibnamefont{and} \bibinfo{author}{\bibfnamefont{W.}~\bibnamefont{Ketterle}},
  \bibinfo{journal}{Nature} \textbf{\bibinfo{volume}{392}},
  \bibinfo{pages}{151} (\bibinfo{year}{1998}).

\bibitem[{\citenamefont{Courteille et~al.}(1998)\citenamefont{Courteille,
  Freeland, Heinzen, van Abeelen, and Verhaar}}]{cour98fesh}
\bibinfo{author}{\bibfnamefont{P.}~\bibnamefont{Courteille}},
  \bibinfo{author}{\bibfnamefont{R.~S.} \bibnamefont{Freeland}},
  \bibinfo{author}{\bibfnamefont{D.~J.} \bibnamefont{Heinzen}},
  \bibinfo{author}{\bibfnamefont{F.~A.} \bibnamefont{van Abeelen}},
  \bibnamefont{and} \bibinfo{author}{\bibfnamefont{B.~J.}
  \bibnamefont{Verhaar}}, \bibinfo{journal}{Phys. Rev. Lett.}
  \textbf{\bibinfo{volume}{81}}, \bibinfo{pages}{69} (\bibinfo{year}{1998}).

\bibitem[{\citenamefont{Greiner et~al.}(2003)\citenamefont{Greiner, Regal, and
  Jin}}]{grei03mol_bec}
\bibinfo{author}{\bibfnamefont{M.}~\bibnamefont{Greiner}},
  \bibinfo{author}{\bibfnamefont{C.~A.} \bibnamefont{Regal}}, \bibnamefont{and}
  \bibinfo{author}{\bibfnamefont{D.~S.} \bibnamefont{Jin}},
  \bibinfo{journal}{Nature} \textbf{\bibinfo{volume}{426}},
  \bibinfo{pages}{537} (\bibinfo{year}{2003}).

\bibitem[{\citenamefont{Jochim et~al.}(2003)\citenamefont{Jochim, Bartenstein,
  Altmeyer, Hendl, Riedl, Chin, Denschlag, and Grimm}}]{joch03bec}
\bibinfo{author}{\bibfnamefont{S.}~\bibnamefont{Jochim}},
  \bibinfo{author}{\bibfnamefont{M.}~\bibnamefont{Bartenstein}},
  \bibinfo{author}{\bibfnamefont{A.}~\bibnamefont{Altmeyer}},
  \bibinfo{author}{\bibfnamefont{G.}~\bibnamefont{Hendl}},
  \bibinfo{author}{\bibfnamefont{S.}~\bibnamefont{Riedl}},
  \bibinfo{author}{\bibfnamefont{C.}~\bibnamefont{Chin}},
  \bibinfo{author}{\bibfnamefont{J.~H.} \bibnamefont{Denschlag}},
  \bibnamefont{and} \bibinfo{author}{\bibfnamefont{R.}~\bibnamefont{Grimm}},
  \bibinfo{journal}{Science} \textbf{\bibinfo{volume}{302}},
  \bibinfo{pages}{2101} (\bibinfo{year}{2003}).

\bibitem[{\citenamefont{Zwierlein et~al.}(2003)\citenamefont{Zwierlein, Stan,
  Schunck, Raupach, Gupta, Hadzibabic, and Ketterle}}]{zwie03molBEC}
\bibinfo{author}{\bibfnamefont{M.~W.} \bibnamefont{Zwierlein}},
  \bibinfo{author}{\bibfnamefont{C.~A.} \bibnamefont{Stan}},
  \bibinfo{author}{\bibfnamefont{C.~H.} \bibnamefont{Schunck}},
  \bibinfo{author}{\bibfnamefont{S.~M.~F.} \bibnamefont{Raupach}},
  \bibinfo{author}{\bibfnamefont{S.}~\bibnamefont{Gupta}},
  \bibinfo{author}{\bibfnamefont{Z.}~\bibnamefont{Hadzibabic}},
  \bibnamefont{and} \bibinfo{author}{\bibfnamefont{W.}~\bibnamefont{Ketterle}},
  \bibinfo{journal}{Phys. Rev. Lett.} \textbf{\bibinfo{volume}{91}},
  \bibinfo{pages}{250401} (\bibinfo{year}{2003}).

\bibitem[{\citenamefont{Bourdel et~al.}(2004)\citenamefont{Bourdel, Khaykovich,
  Cubizolles, Zhang, Chevy, Teichmann, Tarruell, Kokkelmans, and
  Salomon}}]{bour04coll}
\bibinfo{author}{\bibfnamefont{T.}~\bibnamefont{Bourdel}},
  \bibinfo{author}{\bibfnamefont{L.}~\bibnamefont{Khaykovich}},
  \bibinfo{author}{\bibfnamefont{J.}~\bibnamefont{Cubizolles}},
  \bibinfo{author}{\bibfnamefont{J.}~\bibnamefont{Zhang}},
  \bibinfo{author}{\bibfnamefont{F.}~\bibnamefont{Chevy}},
  \bibinfo{author}{\bibfnamefont{M.}~\bibnamefont{Teichmann}},
  \bibinfo{author}{\bibfnamefont{L.}~\bibnamefont{Tarruell}},
  \bibinfo{author}{\bibfnamefont{S.~J. J. M.~F.} \bibnamefont{Kokkelmans}},
  \bibnamefont{and} \bibinfo{author}{\bibfnamefont{C.}~\bibnamefont{Salomon}},
  \bibinfo{journal}{Phys. Rev. Lett.} \textbf{\bibinfo{volume}{93}},
  \bibinfo{pages}{050401} (\bibinfo{year}{2004}).

\bibitem[{\citenamefont{Regal et~al.}(2004)\citenamefont{Regal, Greiner, and
  Jin}}]{rega04}
\bibinfo{author}{\bibfnamefont{C.~A.} \bibnamefont{Regal}},
  \bibinfo{author}{\bibfnamefont{M.}~\bibnamefont{Greiner}}, \bibnamefont{and}
  \bibinfo{author}{\bibfnamefont{D.~S.} \bibnamefont{Jin}},
  \bibinfo{journal}{Phys. Rev. Lett.} \textbf{\bibinfo{volume}{92}},
  \bibinfo{pages}{040403} (\bibinfo{year}{2004}).

\bibitem[{\citenamefont{Zwierlein et~al.}(2004)\citenamefont{Zwierlein, Stan,
  Schunck, Raupach, Kerman, and Ketterle}}]{zwie04rescond}
\bibinfo{author}{\bibfnamefont{M.~W.} \bibnamefont{Zwierlein}},
  \bibinfo{author}{\bibfnamefont{C.~A.} \bibnamefont{Stan}},
  \bibinfo{author}{\bibfnamefont{C.~H.} \bibnamefont{Schunck}},
  \bibinfo{author}{\bibfnamefont{S.~M.~F.} \bibnamefont{Raupach}},
  \bibinfo{author}{\bibfnamefont{A.~J.} \bibnamefont{Kerman}},
  \bibnamefont{and} \bibinfo{author}{\bibfnamefont{W.}~\bibnamefont{Ketterle}},
  \bibinfo{journal}{Phys. Rev. Lett.} \textbf{\bibinfo{volume}{92}},
  \bibinfo{pages}{120403} (\bibinfo{year}{2004}).
  
\bibitem[{\citenamefont{Kinast et~al.}(2004{\natexlab{a}})\citenamefont{Kinast,
  Hemmer, Gehm, Turlapov, and Thomas}}]{kina04sfluid}
\bibinfo{author}{\bibfnamefont{J.}~\bibnamefont{Kinast}},
  \bibinfo{author}{\bibfnamefont{S.~L.} \bibnamefont{Hemmer}},
  \bibinfo{author}{\bibfnamefont{M.~E.} \bibnamefont{Gehm}},
  \bibinfo{author}{\bibfnamefont{A.}~\bibnamefont{Turlapov}}, \bibnamefont{and}
  \bibinfo{author}{\bibfnamefont{J.~E.} \bibnamefont{Thomas}},
  \bibinfo{journal}{Phys. Rev. Lett.} \textbf{\bibinfo{volume}{92}},
  \bibinfo{pages}{150402} (\bibinfo{year}{2004}{\natexlab{a}}).

\bibitem[{\citenamefont{Eagles}(1969)}]{eagl69}
\bibinfo{author}{\bibfnamefont{D.~M.} \bibnamefont{Eagles}},
  \bibinfo{journal}{Phys. Rev.} \textbf{\bibinfo{volume}{186}},
  \bibinfo{pages}{456–463} (\bibinfo{year}{1969}).

\bibitem[{\citenamefont{Leggett}(1980)}]{legg80}
\bibinfo{author}{\bibfnamefont{A.~J.} \bibnamefont{Leggett}}, in
  \emph{\bibinfo{booktitle}{Modern Trends in the Theory of Condensed Matter.
  Proceedings of the XVIth Karpacz Winter School of Theoretical Physics,
  Karpacz, Poland, 1980,}} (\bibinfo{publisher}{Springer-Verlag, Berlin},
  \bibinfo{address}{Karpacz, Poland}, \bibinfo{year}{1980}), pp.
  \bibinfo{pages}{13--27}.

\bibitem[{\citenamefont{Nozières and Schmitt-Rink}(1985)}]{nozi85}
\bibinfo{author}{\bibfnamefont{P.}~\bibnamefont{Nozières}} \bibnamefont{and}
  \bibinfo{author}{\bibfnamefont{S.}~\bibnamefont{Schmitt-Rink}},
  \bibinfo{journal}{J. Low Temp. Phys.} \textbf{\bibinfo{volume}{59}},
  \bibinfo{pages}{195} (\bibinfo{year}{1985}).

\bibitem[{\citenamefont{Falco and Stoof}(2004)}]{falc04}
\bibinfo{author}{\bibfnamefont{G.~M.} \bibnamefont{Falco}} \bibnamefont{and}
  \bibinfo{author}{\bibfnamefont{H.~T.~C.} \bibnamefont{Stoof}},
  \bibinfo{journal}{Phys. Rev. Lett.} \textbf{\bibinfo{volume}{92}},
  \bibinfo{pages}{130401} (\bibinfo{year}{2004}).

\bibitem[{\citenamefont{Barankov and Levitov}(2004)}]{Bara04Coex}
\bibinfo{author}{\bibfnamefont{R.~A.} \bibnamefont{Barankov}} \bibnamefont{and}
  \bibinfo{author}{\bibfnamefont{L.~S.} \bibnamefont{Levitov}},
  \bibinfo{journal}{Phys. Rev. Lett.} \textbf{\bibinfo{volume}{93}},
  \bibinfo{pages}{130403} (\bibinfo{year}{2004}).

\bibitem[{\citenamefont{Ho}()}]{Ho04proj}
\bibinfo{author}{\bibfnamefont{T.-L.} \bibnamefont{Ho}},
  \bibinfo{note}{cond-mat/0404517}.
  
\bibitem[{\citenamefont{Simonucci et~al.}()\citenamefont{Simonucci, Pieri, and
  Strinati}}]{Simo04}
\bibinfo{author}{\bibfnamefont{S.}~\bibnamefont{Simonucci}},
  \bibinfo{author}{\bibfnamefont{P.}~\bibnamefont{Pieri}}, \bibnamefont{and}
  \bibinfo{author}{\bibfnamefont{G.~C.} \bibnamefont{Strinati}},
  \bibinfo{note}{cond-mat/0407600}.
  
\bibitem[{\citenamefont{Matthews et~al.}(1998)\citenamefont{Matthews, Hall,
  Jin, Ensher, Wieman, Cornell, Dalfovo, Minniti, and Stringari}}]{matt98}
\bibinfo{author}{\bibfnamefont{M.~R.} \bibnamefont{Matthews}},
  \bibinfo{author}{\bibfnamefont{D.~S.} \bibnamefont{Hall}},
  \bibinfo{author}{\bibfnamefont{D.~S.} \bibnamefont{Jin}},
  \bibinfo{author}{\bibfnamefont{J.~R.} \bibnamefont{Ensher}},
  \bibinfo{author}{\bibfnamefont{C.~E.} \bibnamefont{Wieman}},
  \bibinfo{author}{\bibfnamefont{E.~A.} \bibnamefont{Cornell}},
  \bibinfo{author}{\bibfnamefont{F.}~\bibnamefont{Dalfovo}},
  \bibinfo{author}{\bibfnamefont{C.}~\bibnamefont{Minniti}}, \bibnamefont{and}
  \bibinfo{author}{\bibfnamefont{S.}~\bibnamefont{Stringari}},
  \bibinfo{journal}{Phys. Rev. Lett.} \textbf{\bibinfo{volume}{81}},
  \bibinfo{pages}{243} (\bibinfo{year}{1998}).

\bibitem[{\citenamefont{Greiner et~al.}()\citenamefont{Greiner, Regal, and
  Jin}}]{grei04spec}
\bibinfo{author}{\bibfnamefont{M.}~\bibnamefont{Greiner}},
  \bibinfo{author}{\bibfnamefont{C.~A.} \bibnamefont{Regal}}, \bibnamefont{and}
  \bibinfo{author}{\bibfnamefont{D.~S.} \bibnamefont{Jin}},
  \bibinfo{note}{cond-mat/0407381}.
  
\bibitem{foot:workshop}
We presented the main result of the present paper at the KITP workshop in Santa Barbara, May 10-14, 2004.

\bibitem[{\citenamefont{Schunck et~al.}()\citenamefont{Schunck, Zwierlein,
  Stan, Raupach, Ketterle, Simoni, Tiesinga, Williams, and
  Julienne}}]{schu04fesh}
\bibinfo{author}{\bibfnamefont{C.~H.} \bibnamefont{Schunck}},
  \bibinfo{author}{\bibfnamefont{M.~W.} \bibnamefont{Zwierlein}},
  \bibinfo{author}{\bibfnamefont{C.~A.} \bibnamefont{Stan}},
  \bibinfo{author}{\bibfnamefont{S.~M.~F.} \bibnamefont{Raupach}},
  \bibinfo{author}{\bibfnamefont{W.}~\bibnamefont{Ketterle}},
  \bibinfo{author}{\bibfnamefont{A.}~\bibnamefont{Simoni}},
  \bibinfo{author}{\bibfnamefont{E.}~\bibnamefont{Tiesinga}},
  \bibinfo{author}{\bibfnamefont{C.~J.} \bibnamefont{Williams}},
  \bibnamefont{and} \bibinfo{author}{\bibfnamefont{P.~S.}
  \bibnamefont{Julienne}}, \bibinfo{note}{cond-mat/0407373}.

\bibitem[{\citenamefont{Bartenstein et~al.}()\citenamefont{Bartenstein,
  Altmeyer, Riedl, Geursen, Jochim, Chin, Denschlag, Grimm, Simoni, Tiesinga
  et~al.}}]{bart04fesh}
\bibinfo{author}{\bibfnamefont{M.}~\bibnamefont{Bartenstein}},
  \bibinfo{author}{\bibfnamefont{A.}~\bibnamefont{Altmeyer}},
  \bibinfo{author}{\bibfnamefont{S.}~\bibnamefont{Riedl}},
  \bibinfo{author}{\bibfnamefont{R.}~\bibnamefont{Geursen}},
  \bibinfo{author}{\bibfnamefont{S.}~\bibnamefont{Jochim}},
  \bibinfo{author}{\bibfnamefont{C.}~\bibnamefont{Chin}},
  \bibinfo{author}{\bibfnamefont{J.~H.} \bibnamefont{Denschlag}},
  \bibinfo{author}{\bibfnamefont{R.}~\bibnamefont{Grimm}},
  \bibinfo{author}{\bibfnamefont{A.}~\bibnamefont{Simoni}},
  \bibinfo{author}{\bibfnamefont{E.}~\bibnamefont{Tiesinga}},
  \bibinfo{author}{\bibfnamefont{C.~J.}~\bibnamefont{Williams}},
  \bibnamefont{and}
  \bibinfo{author}{\bibfnamefont{P.~S.}~\bibnamefont{Julienne}},
  \bibinfo{note}{cond-mat/0408673}.

\bibitem[{\citenamefont{Carr et~al.}(2004)\citenamefont{Carr, Shlyapnikov, and
  Castin}}]{carr03}
\bibinfo{author}{\bibfnamefont{L.~D.} \bibnamefont{Carr}},
  \bibinfo{author}{\bibfnamefont{G.~V.} \bibnamefont{Shlyapnikov}},
  \bibnamefont{and} \bibinfo{author}{\bibfnamefont{Y.}~\bibnamefont{Castin}},
  \bibinfo{journal}{Phys. Rev. Lett.} \textbf{\bibinfo{volume}{92}},
  \bibinfo{pages}{150404} (\bibinfo{year}{2004}).
  
\bibitem{foot:fieldcalib}
The instantaneous magnetic field was determined by probing the atoms using a Zeeman-sensitive optical transition. For the 500 Hz modulation, the deduced field followed the modulation current with a time delay of $(85 \pm 5)\, \mu \rm s$ and a reduced amplitude of 95\% compared to
the dc situation. This was due to induced eddy currents in the apparatus.

\bibitem{foot:noatoms}
The transfer probability depends on the ramp speed and on the density of the cloud. In a
tighter trap with 150 mW of power we cannot discern any unpaired atoms after the ramp.

\bibitem{foot:evaporation}
This was far shorter than
evaporation timescales, which were on the order of 100 ms.

\bibitem[{\citenamefont{Barankov et~al.}(2004)\citenamefont{Barankov, Levitov,
  and Spivak}}]{bara04Rabi}
\bibinfo{author}{\bibfnamefont{R.~A.} \bibnamefont{Barankov}},
  \bibinfo{author}{\bibfnamefont{L.~S.} \bibnamefont{Levitov}},
  \bibnamefont{and} \bibinfo{author}{\bibfnamefont{B.~Z.}
  \bibnamefont{Spivak}}, \bibinfo{journal}{Phys. Rev. Lett.}
  \textbf{\bibinfo{volume}{93}}, \bibinfo{pages}{160401}
  (\bibinfo{year}{2004}).

\bibitem[{\citenamefont{Bartenstein et~al.}(2004)\citenamefont{Bartenstein,
  Altmeyer, Riedl, Jochim, Chin, Denschlag, and Grimm}}]{bart04coll}
\bibinfo{author}{\bibfnamefont{M.}~\bibnamefont{Bartenstein}},
  \bibinfo{author}{\bibfnamefont{A.}~\bibnamefont{Altmeyer}},
  \bibinfo{author}{\bibfnamefont{S.}~\bibnamefont{Riedl}},
  \bibinfo{author}{\bibfnamefont{S.}~\bibnamefont{Jochim}},
  \bibinfo{author}{\bibfnamefont{C.}~\bibnamefont{Chin}},
  \bibinfo{author}{\bibfnamefont{J.~H.} \bibnamefont{Denschlag}},
  \bibnamefont{and} \bibinfo{author}{\bibfnamefont{R.}~\bibnamefont{Grimm}},
  \bibinfo{journal}{Phys. Rev. Lett.} \textbf{\bibinfo{volume}{92}},
  \bibinfo{pages}{203201} (\bibinfo{year}{2004}).

\bibitem[{\citenamefont{Kinast et~al.}(2004{\natexlab{b}})\citenamefont{Kinast,
  Turlapov, and Thomas}}]{kina04hydr}
\bibinfo{author}{\bibfnamefont{J.}~\bibnamefont{Kinast}},
  \bibinfo{author}{\bibfnamefont{A.}~\bibnamefont{Turlapov}}, \bibnamefont{and}
  \bibinfo{author}{\bibfnamefont{J.~E.} \bibnamefont{Thomas}},
  \bibinfo{journal}{Phys. Rev. A} \textbf{\bibinfo{volume}{70}},
  \bibinfo{pages}{051401(R)} (\bibinfo{year}{2004}{\natexlab{b}}).
  
\bibitem{foot:pure}
Indeed, on resonance we observe almost pure condensates, and only a negligible amount of unpaired atoms after the ramp. Note that fitting routines,
saturated absorption and imaging noise all tend to underestimate
condensate fractions.
    
\bibitem{foot:overestimate}
This is not true in the BCS regime, where the atomic density is independent of the presence of a condensate. Still, an additional overestimation of the condensate fraction comes from the fact that the condensed pairs are concentrated
in the high density region of the cloud, where the conversion efficiency is higher. However, this does not affect the total molecular signal.


\end{thebibliography}
\end{document}